\documentclass[fleqn,usenatbib]{mnras}

\usepackage{newtxtext,newtxmath}

\usepackage[T1]{fontenc}

\DeclareRobustCommand{\VAN}[3]{#2}
\let\VANthebibliography\thebibliography
\def\thebibliography{\DeclareRobustCommand{\VAN}[3]{##3}\VANthebibliography}

\usepackage{graphicx}	
\usepackage{amsmath}	
\usepackage{subcaption} 
\title{The slower rotation of planet-hosting stars revisited}

\author[A. Binnenfeld et al.]{
Avraham Binnenfeld,$^{1}$\thanks{E-mail: avrahambinn@gmail.com}
Amitay Sussholz,$^{2}$
Simchon Faigler,$^{2}$
and Tsevi Mazeh$^{2,3}$
\\
$^{1}$ Institute of Physics, École Polytechnique Fédérale de Lausanne (EPFL), Observatoire de Sauverny, Chemin Pegasi 51b, 1290 Versoix, Switzerland\\
$^{2}$School of Physics and Astronomy, Tel Aviv University, Tel Aviv 6997801, Israel\\
$^{3}$Max-Planck-Institut für Astronomie, Königstuhl 17, D-69117 Heidelberg, Germany
}

\date{Accepted XXX. Received YYY; in original form ZZZ}

\pubyear{\the\year{}}

\begin{document}
\label{firstpage}
\pagerange{\pageref{firstpage}--\pageref{lastpage}}
\maketitle

\begin{abstract}
%
Previous studies combined rotation periods derived from \textit{Kepler} photometry, stellar parameters from previous {\it Gaia} Data Releases, and planet hosts from the NASA Exoplanet Archive to conclude that statistically planet-hosting stars rotate slower than stars without detected planets. 
Here, we revisit these studies by using similar {\it Kepler} rotational period sample, 
but 
with an updated {\it Gaia} DR3-based stellar parameter catalogue and an updated exoplanet catalogue.
We apply minimal sample cuts in evolutionary state, metallicity, and distance, and then model the stellar rotation period as a function of effective temperature alone. 
For each star, we derive the ratio of its observed period to the temperature-expected period, and compare the ratio distributions for the planet-host and non-host stars.
For periods shorter than 20 d, the planet-host distribution of this ratio is shifted towards longer periods; the period-ratio median is $\sim15\%$ larger for hosts than for non-hosts ($p_{\rm KS}=2.4\times10^{-4}$).
We also examine the recently published TESS All-Sky Rotation Survey (TARS) and find a similar trend.
However, the strongly structured period distribution precludes a robust comparison between the host and the non-host stars.
Therefore, the updated 
analysis, using more robust stellar and planetary parameters, recovers the previously reported association, strengthening the case for a genuine connection between stellar rotation and planet occurrence.
\end{abstract}




\section{Introduction}
\label{sec:introduction}

Rotation is a key aspect of stellar angular-momentum (AM) evolution. It is shaped through pre-main-sequence stages by disc interactions and subsequent magnetic braking, and is strongly dependent on stellar age and effective temperature \citep[e.g.,][]{Skumanich1972, 2014prpl.conf..433B,Matt2015}. A relationship between stellar rotation and planetary systems is then expected, as protoplanetary discs play a central role in redistributing AM during star and planet formation \citep[e.g.,][]{Armitage2011,Bai2026}.

Several earlier studies have compared the stellar rotation of planet-hosting stars with that of stars without detected planets, using the samples available at the time \citep[e.g.,][]{Pont2009, Lanza2010, Gonzalez2015,Ceillier2016}.  The \textit{Kepler} Space Telescope mission \citep{Koch2010, Borucki2010} is particularly valuable for studying stellar rotation, as it produced roughly two hundred thousand long, high-precision light curves, enabling robust measurements of photometric periodic modulation. \citet{Sibony2022} presented a detailed statistical comparison of \textit{Kepler} planet hosts with a control sample of stars without detected planets, based on \citet*[][hereafter MMA]{McQuillan2013b,McQuillan2014}. They found that planet-hosting stars rotate, on average, $1.63\pm0.40$\,d slower than their matched controls.
The MMA sample combines the rotation periods measured by \citet{McQuillan2013b} for \textit{Kepler} objects of interest with those measured by \citet{McQuillan2014} for the wider \textit{Kepler} sample.  Both catalogues used the autocorrelation-function (ACF) method. 


In this work, we revisit this result using the same MMA rotation-period catalogue. Retaining the original rotation sample allows for a direct comparison with \citet{Sibony2022}, while updating the stellar parameters, planet-host classification, and statistical analysis.
We replace the stellar parameters used in the original analysis with homogeneous measurements from {\it Gaia} DR3 \citep{Gaia2023Vallenari, Andrae2023}, and update the planet-host sample using the current exoplanet census. 
We model the dependence of rotation period on effective temperature alone and derive for each star the ratio of its observed period to the temperature-expected period.  We then compare the ratio distributions for the planet-host and non-host stars, 
rather than characterizing the result primarily through a difference between median periods.
We use a fractional measure because the same difference in days can represent a very different effect for fast and slow rotators. This allows us to compare the host--non-host offset relative to the typical rotation period.
With updated data and a straightforward analysis, we reproduce the signal reported by \citet{Sibony2022}: stars with detected planets are systematically shifted towards longer rotation periods. 

We also examine the TESS \citep{Ricker2015} All-Sky Rotation Survey \citep[TARS;][]{Boyle2026TARS} as a possible independent test. However, the inferred host--non-host difference is sensitive to sample selection, and the strongly structured TARS period distribution prevents such a comparison. 

The paper is organized as follows. In Section~\ref{sec:kepler}, we present the updated analysis of the \textit{Kepler} sample. In Section~\ref{sec:tess}, we perform an analogous host--non-host comparison using TARS and assess its robustness. In Section~\ref{sec:discussion}, we summarize the results and discuss their observational and astrophysical implications.

\section{An updated \textit{Kepler} analysis}
\label{sec:kepler}

We revisit the result of \citet{Sibony2022} using the same underlying MMA rotation-period catalogue, but updated stellar parameters, sample selection, and planet-host classification.
We cross-match the MMA stars by KIC identifier to {\it Gaia} DR3 \citep{Gaia2023Vallenari}.  Effective temperatures, parallaxes, photometry, extinction estimates, and metallicities are taken from {\it Gaia} DR3 rather than from the catalogue of \citet{Berger2020} used by \citet{Sibony2022}.  We adopt the GSP-Phot parameters \citep{Andrae2023}.

\begin{figure}
\centering
\includegraphics[width=0.9\columnwidth]{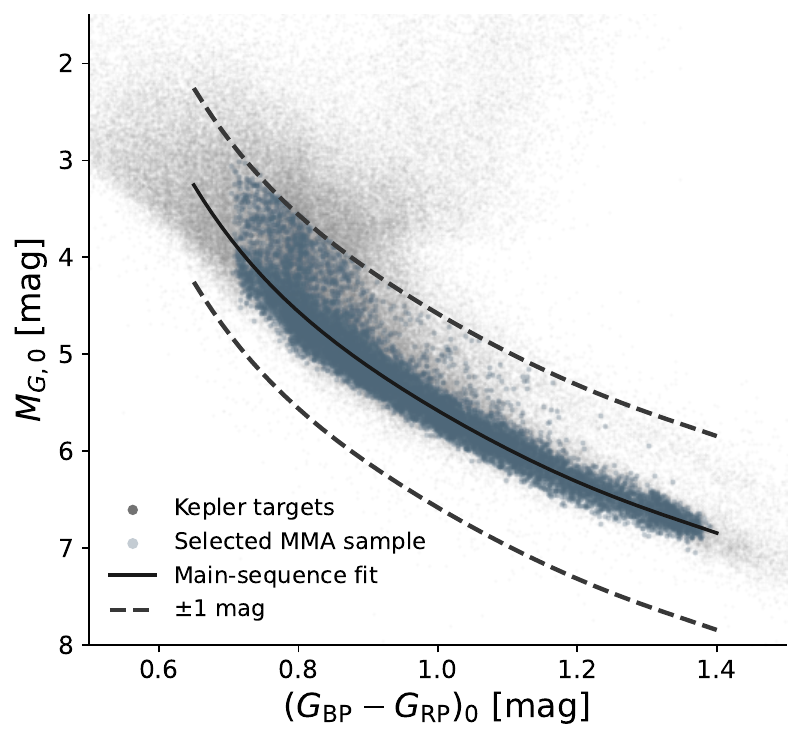}
\caption{Extinction-corrected {\it Gaia} colour--magnitude diagram of \textit{Kepler} targets (grey) and the selected MMA sample (blue). The solid curve shows the fitted main-sequence region. The dashed curves mark the $\pm1$ mag limits.}
\label{fig:kepler_full_cmd_with_mma_selection}
\end{figure}

We restrict the analysis to stars with $4500\leq T_{\mathrm{\tiny eff}}\leq6000\,{\rm K}$. To remove evolved stars, we construct an extinction-corrected {\it Gaia} colour--magnitude diagram and fit the main-sequence region (see Fig.~\ref{fig:kepler_full_cmd_with_mma_selection}). We retain stars within 1 mag of the fitted region. Detected planet hosts are stars associated with planets in the NASA Exoplanet Archive, accessed on 2026 September 4 \citep{Christiansen2025}. All remaining selected MMA stars are used as non-hosts. Throughout the paper, ``non-hosts'' therefore means stars with no detected planets.

\begin{figure}
\centering
\includegraphics[width=0.9\columnwidth]{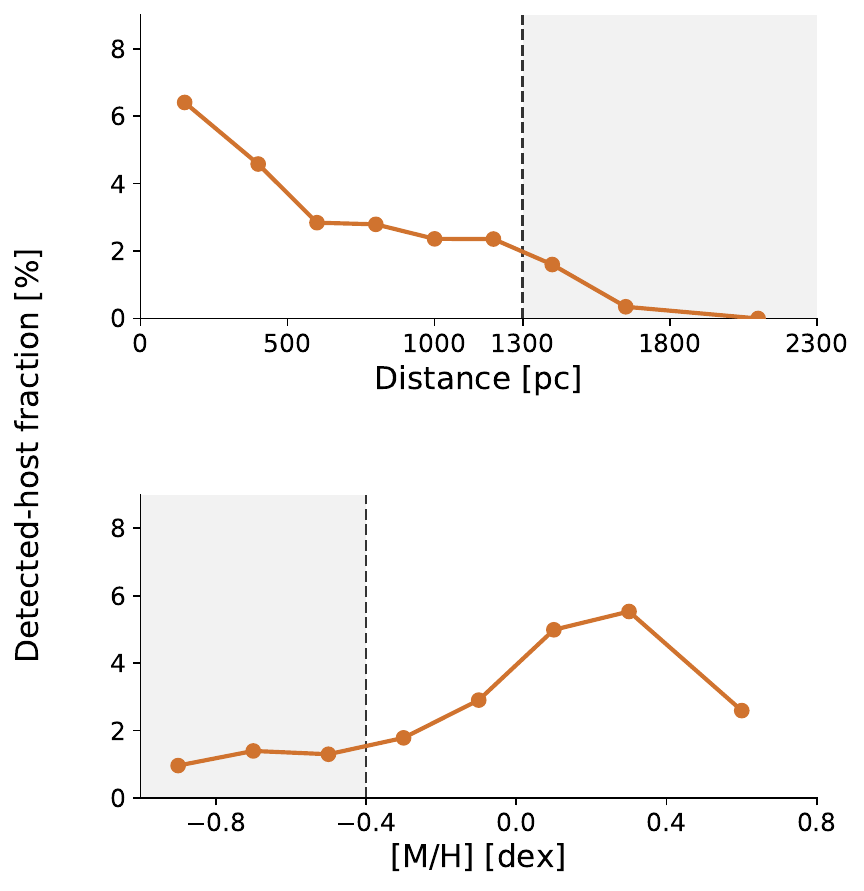}
\caption{Detected-host fraction as a function of distance (top) and {\it Gaia} DR3 metallicity (bottom). The distance panel includes stars with $[\mathrm{M/H}]>-0.4$, while the metallicity panel includes stars within 1300 pc. The grey regions mark the ranges excluded from the final sample.}
\label{fig:kepler-selection-support}
\end{figure}

We further restrict the comparison to the region of distance and metallicity in which detected hosts are sufficiently represented.
We adopt $d\leq1300$ pc and $[\mathrm{M/H}]_{\mathrm{\tiny GSP-Phot}}>-0.4$. 
As shown in Fig.~\ref{fig:kepler-selection-support}, the detected-host fraction drops sharply beyond these limits. The following results are not sensitive to small changes in these boundaries.
Before applying any rotation-period cut, the retained region contains 400 detected hosts among 12\,525 stars, or 3.2\,\%. By comparison, only 9 of the 902 otherwise eligible metal-rich stars beyond 1.3 kpc and 18 of the 1\,398 metal-poor stars within 1.3 kpc are detected hosts, corresponding to 1.0 and 1.3\,\%, respectively.

We restrict the analysis to $1<P_{\rm rot}\leq30$ d, adopting a conservative upper limit motivated by the increasing sensitivity of long-period \textit{Kepler} rotation measurements to instrumental systematics and detrending \citep[see][]{Nielsen2013}.

\begin{figure}
\centering
\includegraphics[width=\columnwidth]{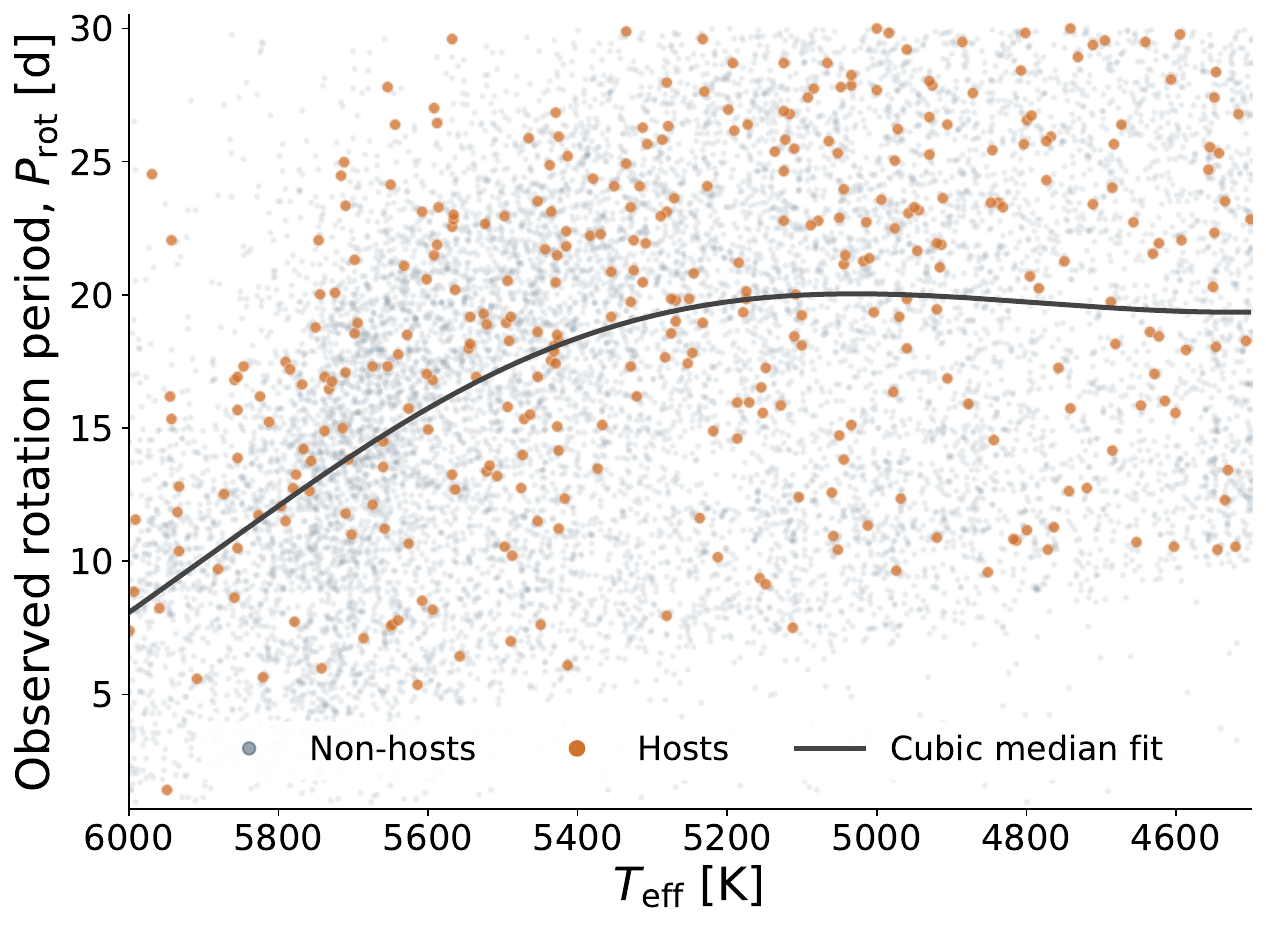}
\caption{Observed rotation period versus {\it Gaia} DR3 effective temperature for the selected \textit{Kepler} sample with $1<P_{\text{\tiny rot}}\leq30$ d. The line is a third-degree polynomial describing the temperature dependence of the non-host rotations.}
\label{fig:kepler-temperature-period}
\end{figure}

\begin{figure*}
\centering
\includegraphics[width=0.89\textwidth]{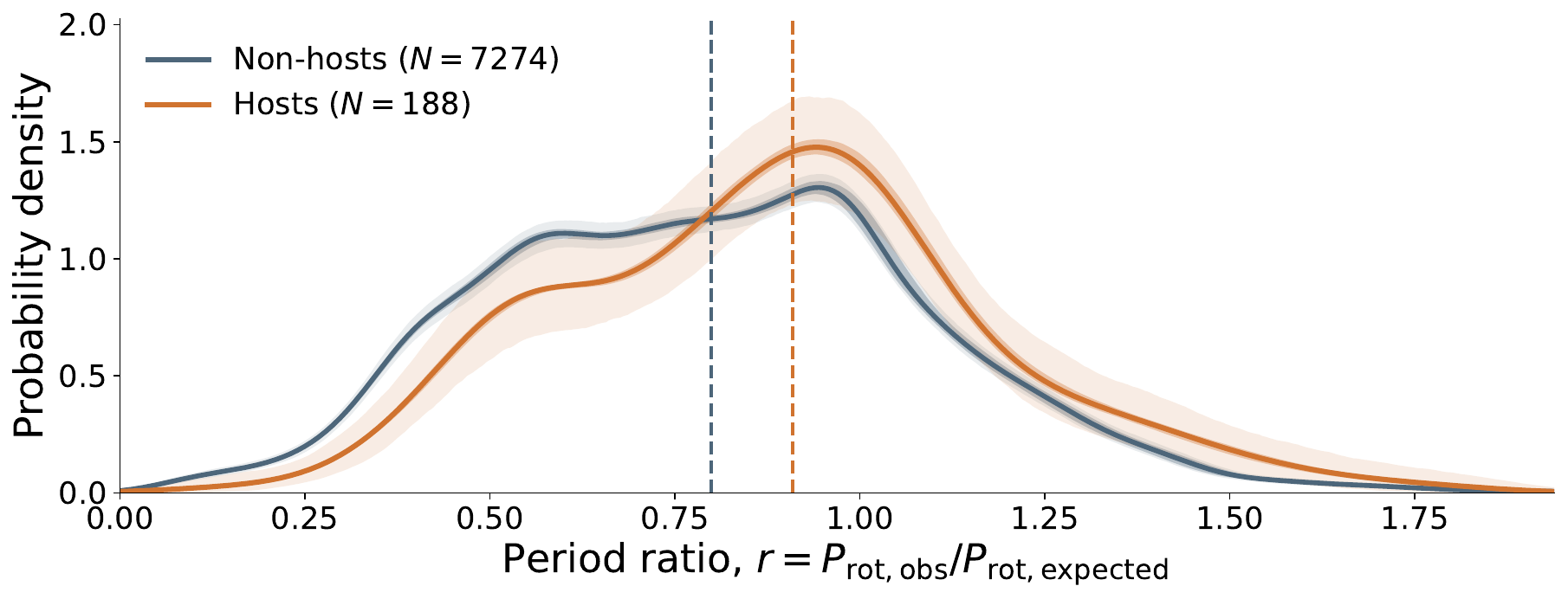}
\caption{Kernel-density estimates of the period ratio for \textit{Kepler} hosts and non-hosts with $1<P_{\rm rot}\leq20$ d. Dashed coloured lines mark the group medians. The darker bands show the $2\sigma$ uncertainty in the temperature fit, while the lighter bands show the combined $2\sigma$ uncertainty in the fit and the KDEs. The median period ratios are 0.909 for hosts and 0.800 for non-hosts, giving a ratio of medians of 1.137.}
\label{fig:kepler-ratio-distributions}
\end{figure*}

Figure~\ref{fig:kepler-temperature-period} shows the temperature--period plane for the selected sample. We fit the median non-host rotation period over $1<P_{\rm rot}\leq30$ d with a third-degree polynomial and use this relation as the temperature reference for all stars. We model this relation as
\begin{equation}
\ln P_{\text{\tiny rot,expected}}=a_0+a_1x+a_2x^2+a_3x^3, \qquad x=\frac{T_{\text{\tiny eff}}-5500\,{\rm K}}{1000\,{\rm K}}.
\label{eq:temperature-fit}
\end{equation}
The coefficients obtained from the full non-host sample are
\begin{equation}
\begin{aligned}
a_0 &= 2.846\pm0.007, &\qquad a_1 &= -0.769\pm0.022,\\
a_2 &= -1.211\pm0.061, &\qquad a_3 &= -0.559\pm0.074.
\end{aligned}
\label{eq:temperature-fit-coefficients}
\end{equation}
The uncertainties are the $1\sigma$ standard deviations obtained from bootstrap resampling of the full non-host sample.

For every star we first define the temperature-corrected period ratio
\begin{equation}
r=\frac{P_{\text{\tiny rot,obs}}}{P_{\text{\tiny rot,expected}}}.
\label{eq:r-definition}
\end{equation}

We then compare the $1<P_{\rm rot}\leq20$ d and $20<P_{\rm rot}\leq30$ d intervals separately. The temperature relation describes the population trend but is not very precise for individual stars. For the non-hosts, the $1\sigma$-equivalent interval of the fitted period ratios is $0.57<r<1.34$, reflecting the substantial scatter around the relation. We also tried adding the {\it Gaia} FLAME age, but as the improvement was negligible we retained the temperature-only correction.

Figure~\ref{fig:kepler-ratio-distributions} shows the $r$ distributions and their two-standard-deviation uncertainty intervals for $1<P_{\text{\tiny rot}}\leq20$ d, a sample containing 188 hosts and 7\,274 non-hosts. 
The median period ratios are 0.909 for hosts and 0.800 for non-hosts, respectively, giving a ratio of medians of 1.14. Therefore, after applying the same temperature correction to both populations, the characteristic rotation period of the hosts is $13.7^{+2.2}_{-5.0}\%$ longer relative to that of the non-hosts. The two-sided KS test gives $D_{\rm KS}=0.156$ and $p_{\rm KS}=2.4\times10^{-4}$.
We find no significant difference between the distributions in the $20<P_{\text{\tiny rot}}\leq30$ d interval.

\begin{figure}
\centering
\includegraphics[width=0.99\columnwidth]{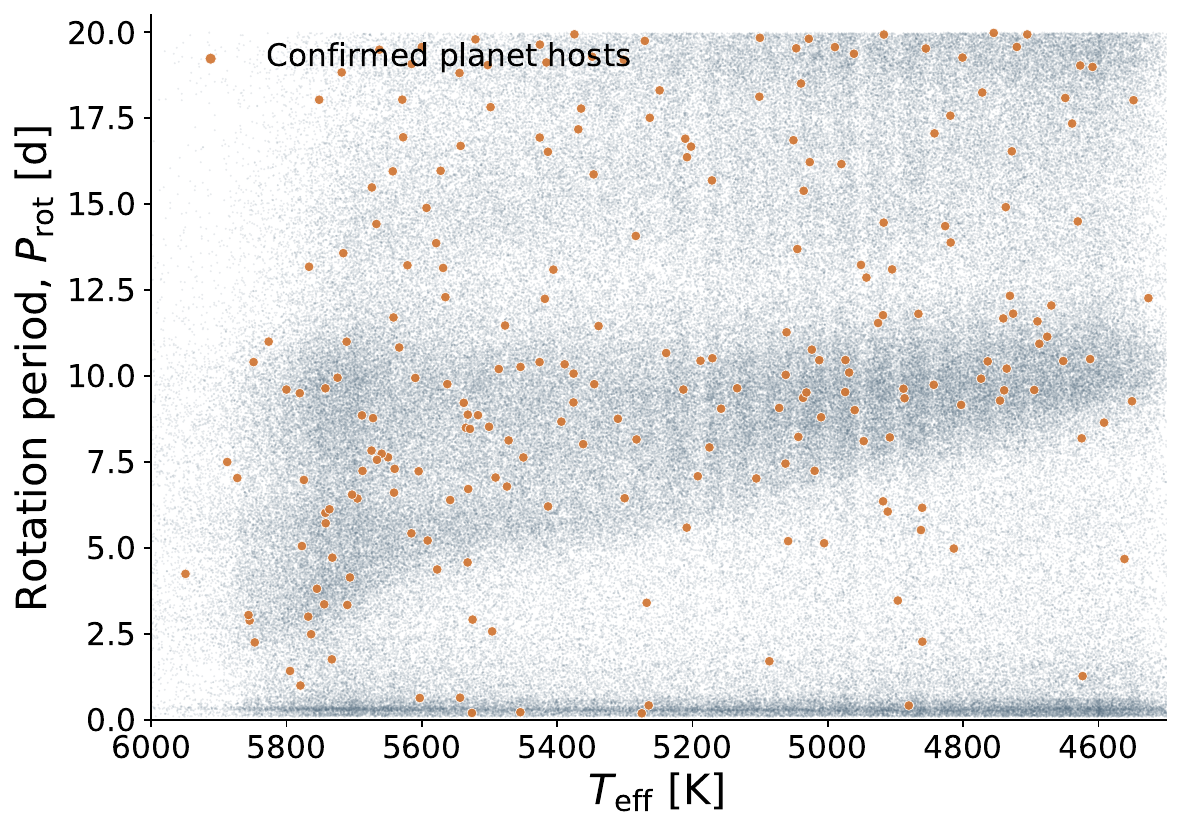}
\caption{TARS rotation periods as a function of effective temperature. Orange points mark the confirmed planet hosts in the sample. Short-period excess and concentration of measurements in several narrow ranges make a straightforward comparison between TARS hosts and non-hosts unreliable.}
\label{fig:tars-period-distribution}
\end{figure}

\section{\textit{TESS} rotational periods}
\label{sec:tess}

Previous work using \textit{TESS} rotation periods has focused mainly on known or candidate planet hosts rather than on a broad host--non-host comparison. \citet{CantoMartins2020} and \citet{Howard2021} measured rotation periods for subsets of TOIs, while \citet{Messias2022} found that the apparent deficit of close-in planets around rapid rotators becomes weaker in a larger \textit{Kepler} and \textit{TESS} sample.

Recently, \citet{Boyle2026TARS} published  a catalogue of stellar rotational periods (TARS) based on the  \textit{TESS} All-Sky Rotation Survey data. 
We assess whether TARS can provide an independent test of the host–non-host rotation difference.
We use version 2 of the TARS catalogue and apply the same {\it Gaia} DR3 temperature, parallax, metallicity, and main-sequence requirements used for the \textit{Kepler} analysis, apart from the distance cut, since TARS is already limited to 500 pc. Hosts are stars with a confirmed planet in the NASA Exoplanet Archive, while planet candidates are excluded from both groups.

For $P_{\text{\tiny rot}}\leq20$ d, the nominal sample contains 228 hosts and 285\,972 non-hosts. The median period ratio is about 8.5\% larger for hosts than for non-hosts, with $p_{\text{\tiny KS}}=5.6\times10^{-3}$, formally in the same direction as the \textit{Kepler} result. However, the result is not stable to reasonable catalogue cuts, and the TARS period distribution itself shows substantial structure (Figure~\ref{fig:tars-period-distribution}), including a strong short-period excess and concentrations at several preferred periods.

Longer-period measurements are also difficult to interpret. TARS reports that its alias-correction procedure can recover periods as long as 25 d from a single sector \citep{Boyle2026TARS}. However, an empirical comparison of K2 and \textit{TESS} periods found that completeness and reliability decline sharply beyond $\sim12$ d due to the 27-d sector baseline, while combining consecutive sectors does not remove persistent systematics \citep{Boyle2025}. These limitations are important when trying to recover a relatively small population-level shift.

We therefore do not regard the current TARS catalogue as sufficiently reliable to test whether the \textit{Kepler} shift is reproduced with \textit{TESS}. A cleaner period distribution and more robust long-period measurements will be required for such a comparison.

\section{Discussion and conclusions}
\label{sec:discussion}

We revisited the reported tendency of detected planet hosts to rotate slower than stars without detected planets, using {\it Gaia} DR3 stellar parameters, an updated planet census, and a simple statistical analysis. In the MMA sample, detected planet hosts are shifted towards longer rotation periods for $1<P_{\text{\tiny rot}}\leq20$ d, with a median temperature-corrected period ratio about 15\% larger than that of the non-hosts ($p_{\text{\tiny KS}}=2.4\times10^{-4}$). No significant difference is found for $20<P_{\text{\tiny rot}}\leq30$ d.

Expressed fractionally, our offset is comparable in scale to the result of \citet{Sibony2022}, who measured a rotation period difference of $1.63\pm0.40$ d. In our sample, the corresponding temperature-corrected median offset is $1.51\pm0.50$ d. They also found that the difference weakened when rapidly rotating stars were excluded and disappeared for $P_{\rm rot}>15$ d,
similar to our null result for periods above 20~d.
This suggests that the apparent effect could be confined to a certain period range, potentially reflecting the strong period dependence of rotation-period detectability.

We examined whether differences in stellar and observational properties could account for the observed host--non-host difference. The hosts are more metal-rich, as expected from the known planet--metallicity correlation \citep[e.g.,][]{FischerValenti2005}, and are slightly brighter, as expected for a sample selected through detection of planetary transits.
However, these differences have almost no effect on the detected period shift, as shown by sample-matching tests.

We note that our comparison is limited by incompleteness in both planet detection and rotation-period recovery. Rapid rotators tend to be more active and variable, making planetary-transit detection more difficult \citep[e.g.,][]{Pont2009,Christiansen2020}, while photometric rotation periods are more easily recovered for stronger and shorter-period modulation \citep[e.g.,][]{McQuillan2014,Masuda2022}. Such biases can contribute to the observed difference, although the quality of the \textit{Kepler} observations minimizes these effects.

Our result is limited to the period range in which the MMA catalogue is comparatively reliable. Slow stellar rotations are harder to measure, and long periods are more susceptible to aliases and catalogue-dependent systematics \citep[e.g.,][]{Nielsen2013,Gonzalez2025}. We therefore concentrate on periods shorter than 20 d. 

We tested the TARS catalogue and find that the nominal sample yields a shift in the same direction as the \textit{Kepler} result. However, the result is not stable to reasonable catalogue cuts, and the TARS period distribution shows substantial structure. We therefore do not regard the current TARS catalogue as a robust independent test.

If the \textit{Kepler} association is indeed astrophysical, several stages of stellar and planetary evolution could contribute. Disc lifetimes and magnetic star-disc coupling affect the early rotation of solar-type stars \citep[e.g.,][]{2014prpl.conf..433B}, absorbing some of its angular momentum, while the same discs provide the conditions for planet formation. At later times, tidal exchange can alter the spin of stars with sufficiently close and massive planets \citep[e.g.,][]{Penev2018}.

Further study could make use of observed stellar line broadening, together with estimates of stellar radii, to infer stellar rotational velocities, following, e.g., the approach presented by \citet{sussholz26}. Such analysis would allow this work to be extended to a substantially wider range of rotational periods, thereby providing a more comprehensive view of the difference between host and non-host stars.

\section*{Acknowledgments}

We thank the anonymous referee for the careful and constructive comments, which led us to revisit key aspects of the analysis and substantially improved our manuscript. 
We are deeply grateful to Ravit Helled for her valuable feedback and comments on the manuscript. This work used data from the European Space Agency (ESA) mission \textit{Gaia}, processed by the \textit{Gaia} Data Processing and Analysis Consortium (DPAC). This research has also used the NASA Exoplanet Archive, which is operated by the California Institute of Technology under contract with NASA through the Exoplanet Exploration Program. This work includes data from the \textit{Kepler} and \textit{TESS} missions. Funding for \textit{Kepler} and \textit{TESS} is provided by NASA. 
We used the following code packages for the analysis: SciPy \citep{2020SciPy-NMeth}, PyAstronomy \citep{pya}, and SPARTA \citep{SPARTA2020}.

\section*{Data Availability}

The data underlying this article will be shared on reasonable request to the corresponding author.

\bibliographystyle{mnras}
\bibliography{main}

\label{lastpage}
\end{document}